\documentclass[aps,prb,reprint,showpacs,superscriptaddress,fleqn]{revtex4-1}

\usepackage{graphicx}
\usepackage{amssymb}
\usepackage{amsmath}

\begin{document}

\title{Valley-dependent resonant tunneling through double magnetic barriers in suspended graphene}

\date{\today}

\author{Nojoon Myoung}
\thanks{Present address: Department of Material Science and Engineering, University of
Ioannina, 45110 Ioannina, Greece} \affiliation{Department of
Physics, Chungnam National University, Daejeon 305-764, Republic of
Korea} \affiliation{Institute of Quantum Systems, Chungnam National
University, Daejeon 305-764, Republic of Korea}
\author{Gukhyung Ihm}
\email[Corresponding author: ]{ghihm@cnu.ac.kr}
\affiliation{Department of Physics, Chungnam National University,
Daejeon 305-764, Republic of Korea}

\begin{abstract}
We theoretically investigate the effects of strain-induced
pseudomagnetic fields on the transmission probability and the
ballistic conductance for Dirac fermion transport in suspended
graphene. We show that resonant tunneling through double magnetic
barriers can be tuned by strain in the suspended region. The
valley-resolved transmission peaks are apparently distinguishable
owing to the sharpness of the resonant tunneling. With the specific
strain, the resonant tunneling is completely suppressed for Dirac
fermions occupying the one valley, but the resonant tunneling exists
for the other valley. The valley-filtering effect is expected to be
measurable by strain engineering. The proposed system can be used to
fabricate a graphene valley filter with the large valley
polarization almost 100\%.
\end{abstract}



\maketitle

\section{Introduction}

Graphene, a two-dimensional honeycomb crystal consisting of only
carbon atoms, has received a lot of attention from researchers
during the last several years. The electronic structure of graphene
can be described by linear energy dispersions at six corners of the
Brillouin zone, which are called Dirac points.\cite{R1} Charge
carriers (electrons and holes) near the Dirac points are
characterized by two degenerate valleys in addition to the spin
degree of freedom, leading to the four-fold degeneracy. The
existence of the valley degeneracy makes graphene a potential
candidate as a new class of nanoelectronic devices,
``valleytronics''.

There have been considerable efforts in the design and realization
of manipulation of the valley index of graphene through diverse
theoretical schemes.\cite{R2,R3,R4,R5,R6,R7,R8,R9,R10} One might use
the fact that valley polarization is produced by nonequilibrium
population of different valleys in a nanoconstriction with zigzag
edges.\cite{R2,R3} A graphene valley filter was suggested by using
line defects.\cite{R4,R5} It is difficult to realize these valley
filters because the atomic-scaled precision of the edges and defects
is required to control the valley transport for practical studies.
On the other hand, some works have focussed on the effects of
substrate-induced strain in graphene in order to construct a
feasible valley filter.\cite{R6,R7,R8,R9,R10} The pseudomagnetic
field indeed plays a role of creating valley-dependent tunneling
with high valley polarization. Despite of the high valley
polarization, one practical issue has remained for the
strain-induced valley filter: the strain induced by substrate is not
controllable. Thus, it is natural to ask whether the tunable
valley-dependent transport is possible.

Research on suspended graphene was originally motivated by the fact
that the advantage of using suspended graphene lies in its ability
to avoid the substrate-induced effects which deteriorate the
electrical properties of graphene.\cite{R11,R12,R13,R14,R15,R16} It
has been demonstrated that suspended graphene sheets show extremely
high mobility due to reduced interactions between graphene sheets
and substrates.\cite{R17,R18} The other significance of suspended
graphene is a potential ability to study the effects of strain on
Dirac fermion transport in graphene. Since graphene has an advantage
of being able to stand elastic deformation up to 20\%,\cite{R19,R20}
the physics of interplay between strain and electronic properties of
graphene becomes a very promising field in recent condensed matter
research. For suspended graphene, the magnitude of strain in a
suspended region can be tuned by controlling the back gate voltage
owing to the Coulomb interaction between a two-dimensional Dirac
fermion gas and a back gate.\cite{R21} The tunable strain of
suspended graphene may offer the opportunity to study how one can
control the transport of Dirac fermions by mechanical strain.

\begin{figure}
\includegraphics[width=8.5cm]{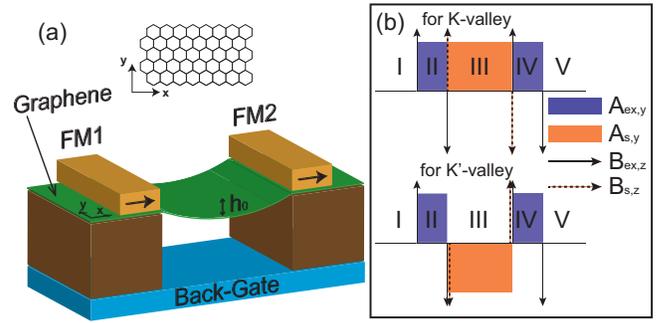}
\caption{(a) Schematics of the model considered in this study. A
graphene sheet is suspended between two clamps, and strain in the
suspended region is induced by an electrostatic pressure via a back
gate. Two FM top gates deposited on unsuspended regions of the
graphene sheet produce a double magnetic barrier structure. The
suspended graphene is placed as its zigzag direction is along the
x-axis of the reference frame in this study. (b) Profiles two kinds
of magnetic fields and vector potentials ($B_{s,z}$ and $B_{ex,z}$;
$A_{s,y}$ and $A_{ex,y}$) for Dirac fermions in different
valleys.}\label{fg:model}
\end{figure}

In this paper, we present a strategy to realize a tunable valley
transport by controlling strain for suspended graphene. We
investigate the valley dependence of the transmission probability
and ballistic conductance because of the existence of pseudomagnetic
fields, with the minimized deterioration of transport properties
such as carrier mobility or mean free path. We find that suspended
graphene with a double magnetic barrier (DMB) structure can be a
efficient valley filter, and the considerable valley polarization
can be achieved by applying strain to the suspended region. The DMB
structure is produced by two ferromagnetic (FM) top gates placed at
both sides of the suspended region over a trench, as illustrated in
Fig. {\ref{fg:model}. The system comprises of two distinct
components, i.e., pseudo- and external magnetic fields induced by
the strain and the FM gates, respectively. We further show that the
valley-filtering effect is also observed in the ballistic
conductance as a measurable quantity for practical investigations.
The results may provide a way of manipulating the valley degree of
freedom for future valleytronics.

\section{Theoretical background and model}

We begin with the Hamiltonian of a two-dimensional Dirac fermion
gas. The Dirac Hamiltonian with the formation of mechanical strain
in the presence of external electric and magnetic fields is given
by\cite{R22,R23}
\begin{align}
H=&v_{F}\tau_{z}\otimes\vec{\sigma}\cdot\left(\vec{p}+e\vec{A}_{ex}+e\vec{A}_{s}\right)\nonumber\\
&\qquad\qquad\qquad\qquad+\tau_{z}\otimes\sigma_{z}\left(U_{ex}+U_{s}\right),
\end{align}
where $\vec{\sigma}=\left(\sigma_{1},\sigma_{2}\right)$ and
$\tau_{z}=\sigma_{3}$ are Pauli matrices, acting on the sublattice
and valley spaces of graphene, recpectively. In this study, we
assume that the external electric potential is constant as zero for
simplicity. The envelop function corresponding to the above
Hamiltonain is
$\left(\psi_{A,K},\psi_{B,K},\psi_{A,K'},\psi_{B,K'}\right)^{T}$.
Both the in- and out-of-plane stresses induce effective vector and
scalar potentials as below:
\begin{align}
\left(\begin{array}{c}A_{s,x}\\A_{s,y}\end{array}\right)=&\xi\frac{\hbar\beta}{ea_{0}}\left(\begin{array}{cc}\cos{3\Omega}&\sin{3\Omega}\\-\sin{3\Omega}&\cos{3\Omega}\end{array}\right)\left(\begin{array}{c}u_{xx}-u_{yy}\\-2u_{xy}\end{array}\right),\nonumber\\
U_{s}=&U_{0}\left(u_{xx}+u_{yy}\right),
\end{align}
where $a_{0}\approx1.4~{\AA}$ is the carbon-carbon bond length,
$\xi=\pm1$ for different valleys,
$\beta=-C\left(\partial\log{t}\right)/\left(\partial\log{a_{0}}\right)\approx2$
with $\mathcal{O}\left(C\right)\sim1$, and $U_{0}=10~eV$. Note that
we set $\Omega=0$ or $\pi/2$ when the strain is applied along the
armchair or zigzag directions. The strain tensor $u_{ij}$ is derived
as follows:
\begin{align}
u_{ij}=\frac{1}{2}\left(\frac{\partial u_{i}}{\partial
r_{j}}+\frac{\partial u_{j}}{\partial
r_{i}}\right)+\frac{1}{2}\frac{\partial h}{\partial
r_{i}}\frac{\partial h}{\partial r_{j}},
\end{align}
where $\vec{u}\left(\vec{r}\right)$ and $h\left(\vec{r}\right)$ are
in- and out-of-plane displacements, respectively.

It has been shown that the homogeneous effective gauge field is
generated in the suspended region when the profile of vertical
deformation of suspended graphene sheet is parabolic.\cite{R21}
Throughout this paper, we assume that the suspended region of
graphene sheet undergoes the parabolic vertical deformation,
allowing convenience of the theoretical framework. In this case, we
have the following profile of vertical deformation:
\begin{align}
h\left(x\right)=\frac{4h_{0}}{L^{2}}\left(x^{2}-\frac{L^{2}}{4}\right),
\end{align}
where the maximum deformation $h_{0}$ is given by\cite{R21}
\begin{align}
h_{0}=\left(\frac{3\pi}{64}\frac{e^{2}}{\epsilon
E}n^{2}L^{4}\right)^{\frac{1}{3}},
\end{align}
where $E=22~eV{\AA}^{-2}$ is Young's modulus of graphene, $\epsilon$
is the electric permittivity, and $L$ is the length of the suspended
region. The in-plane stress, on the other hand, is:
\begin{align}
u_{x}\left(x\right)=\frac{8h_{0}^{2}}{3L^{2}}x-\frac{32h_{0}^{2}}{3L^{4}}x^{3},
\end{align}
satisfying boundary conditions, $u_{x}\left(\pm L/2\right)=0$. We
put $u_{y}=0$ since the in-plane strain is applied along the
$x$-direction. When we have the strain lay along the zigzag
direction, i.e., $\Omega=\pi/2$, the effective vector potential
induced by the strain is given by
\begin{align}
\vec{A}_{s}=\xi\frac{\hbar\beta}{ea_{0}}\frac{8h_{0}^{2}}{3L^{2}}\left[\theta\left(x+\frac{L}{2}\right)-\theta\left(x-\frac{L}{2}\right)\right]\hat{y}.
\end{align}
Note that, in case of $\Omega=0$, the $x$-component of the effective
vector potential becomes constant, and its $y$-component vanishes.
The corresponding pseudomagnetic field
$\vec{B}_{s}=\nabla\times\vec{A}_{s}$ is given by a set of
delta-function spikes at the edges of the suspended region. Now, let
us focus on the effects of pseudomagnetic field with justification
that the effective scalar potential does not dominantly affect on
characteristics of quantum transport in suspended
graphene.\cite{R21} In addition to the pseudomagnetic field, two FM
gates produces a series of magnetic barriers:
\begin{align}
\vec{A}_{ex}=&\left\{B_{1}l_{B}\left[\theta\left(x+W_{1}+\frac{L}{2}\right)-\theta\left(x+\frac{L}{2}\right)\right]\right.\nonumber\\
&\left.+B_{2}l_{B}\left[\theta\left(x-\frac{L}{2}\right)-\theta\left(x-W_{2}-\frac{L}{2}\right)\right]\right\}\hat{y},
\end{align}
where $B_{1,2}$ are the magnetic field strengths due to the FM
gates, $l_{B}=\sqrt{\hbar/eB_{0}}$ is the magnetic length, and
$W_{1,2}$ are the magnetic barrier widths. In the present
discussion, we consider a symmetric case ($B_{1}=B_{2}=B_{ex}$ and
$W_{1}=W_{2}=W$) in order to show the valley-dependent transport,
and we may take into account an asymmetric DMB for manipulating the
spin degree of freedom of Dirac fermions.\cite{R24}

Due to the translational invariance in the $y$-direction, the
wavefunctions are written as $\Psi\left(x,y\right)\propto
e^{ik_{y}y}\psi\left(x\right)$, where
$\psi\left(x\right)=\left(\psi_{A},\psi_{B}\right)^{T}$ of which
components represent sublattices of graphene. By solving Dirac
equation, the wavefunctions are described by plane waves
characterized by the constant $k_{y}$ and the longitudinal
wavevectors $k_{x}$, which can be either exponentially damped or
oscillatory. In a given region $j$ denoted in Fig. \ref{fg:model},
the longitudinal wavevectors read
\begin{align}
k_{x,j}=\sqrt{\epsilon^{2}-k_{y,j}^{2}},
\end{align}
where $k_{y,I}=k_{y,V}=k_{y}$, $k_{y,II}=k_{y,IV}=k_{y}+\beta_{ex}$,
and $k_{y,III}=k_{y}+\xi\beta_{s}$. For convenience, we express all
quantities in dimensionless units, i.e., $\epsilon\rightarrow
E/E_{0}$ and $\vec{k}\rightarrow\vec{k}l_{B}$, by means of
characteristic parameters: for the typical strength of magnetic
fields $B_{0}=0.66~mT$, the magnetic length $l_{B}$ and the energy
$E_{0}=\hbar v_{F}/l_{B}$ are $1~\mu m$ and $0.66~meV$,
respectively. Here, we introduce two important dimensionless
quantities corresponding to the external and pseudomagnetic fields,
$\beta_{ex}=B_{ex}/B_{0}$ and $\beta_{s}=\left(8\beta
l_{B}h_{0}^{2}\right)/\left(3a_{0}L^{2}\right)$, respectively. Note
that $\beta_{s}$ can be tuned by strain while $\beta_{ex}$ are given
by the magnetization of FM gates.

In order to investigate Dirac fermion transport through the
structure, we calculate the transmission and reflection probability.
The wavefunctions for different regions are written as a linear
combination of left- and right-going waves:
\begin{align}
\Psi\left(x,y\right)=&e^{ik_{y,j}y}\left[c_{1}e^{ik_{x,j}x}\left(\begin{array}{c}1\\e^{i\varphi_{j}}\end{array}\right)\right.\nonumber\\
&\qquad\qquad\left.+c_{2}e^{-ik_{x,j}x}\left(\begin{array}{c}1\\-e^{-i\varphi_{j}}\end{array}\right)\right],
\end{align}
where $\varphi_{j}=\tan^{-1}{\left(k_{y,j}/k_{x,j}\right)}$ in
region $j$. The coefficients $c_{1,2}$ become the incidence,
reflection, or transmission coefficient according to regions. For
example, $c_{1}=1$ and $c_{2}=r$ in region $I$; $c_{1}=t$ and
$c_{2}=0$ in region $V$. The transmission coefficient is numerically
calculated by matching conditions that wavefunctions have to be
continuous at interfaces, and the transmission probability is
obtained from the absolute square of the transmission coefficient.
The calculated transmission probability is valley-resolved, i.e.,
$T_{\xi}=\left|t_{\xi}\right|^{2}$.

We also study the ballistic conductance of the proposed system. In
the low temperature limit, the valley-resolved conductance is given
by\cite{R25,R26,R27}
\begin{align}
G_{\xi}\left(E_{F}\right)&=\frac{e^{2}L_{y}k_{F}}{\pi
h}\nonumber\\
&\times\int_{-\frac{\pi}{2}}^{+\frac{\pi}{2}}T_{\xi}\left(E_{F},E_{F}\sin{\phi}\right)\cos{\phi}d\phi,
\end{align}
where $L_{y}$ is the width of the sample in the $y$-direction,
$E_{F}$ is the Fermi energy, and $\phi\equiv\varphi_{I}$ is the
incident angle for Dirac fermions. We introduce the valley
polarization of ballistic conductance:
\begin{align}
P_{v}&=\frac{G_{K}-G_{K'}}{G_{K}+G_{K'}},
\end{align}
where $K$ and $K'$ represent different valleys in Brillouin zone of
graphene.

\section{Results and discussion}

In this section, we show that the resonant tunneling through the DMB
structure depends on the valley index owing to strain effects in the
suspended region. The transmission probability through the structure
is investigated for different magnitudes of the pseudomagnetic field
induced by strain. The angle dependence of the resonant tunneling is
also studied, and the valley-resolved conductance through the
structure is presented to show that it is possible to create highly
valley-polarized transport.

We consider a symmetric DMB with $B_{1}=B_{2}$ and $W_{1}=W_{2}$ in
order to focus on effects of strain in the suspended region of
graphene. As discussed in previous literatures,\cite{R28,R29,R30}
quantum transport in graphene is qualitatively understood by the
effective potential that Dirac fermions experience in a real sense.
The effective potential of the proposed system is given by
\begin{align}
U_{eff,\xi}=\left\{\begin{array}{ll}k_{y}^{2},&\mbox{region I and V},\\
\left(k_{y}+\beta_{ex}\right)^{2},&\mbox{region II and IV},\\
\left(k_{y}+\xi\beta_{s}\right)^{2},&\mbox{region
III},\end{array}\right.\label{eq:effpot}
\end{align}
where $k_{y}=\epsilon\sin{\phi}$. It is straightforward that Dirac
fermion transport through the proposed system depends on the
incident angle and Fermi energy. One can clearly see that, for
normal incidence ($\phi=0$), there is no difference between
transport of Dirac fermions in different valleys because the
effective potential in region $III$ is not changed for valleys,
i.e., $U_{eff,K}=U_{eff,K'}=\beta_{s}^{2}$. On the other hand, for
$\phi\neq0$, Eq. (\ref{eq:effpot}) exhibits a valley-dependent
profile of the effective potential, resulting in valley-dependent
transport.

\begin{figure}
\includegraphics[width=8.5cm]{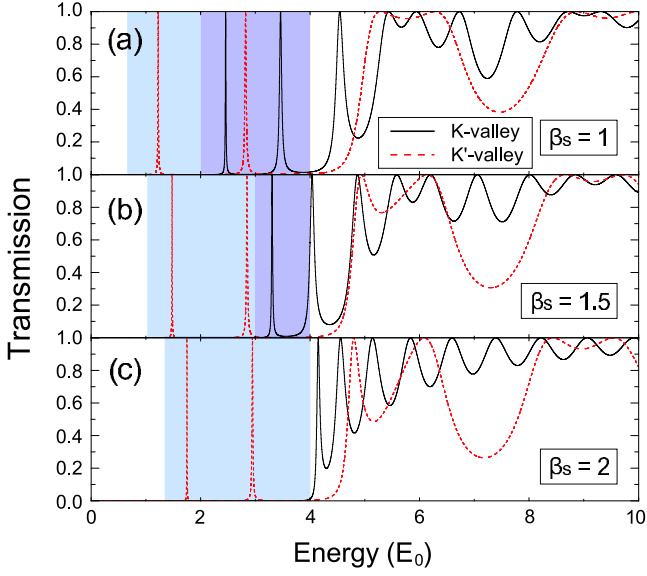}
\caption{ Valley-resolved transmission probability as a function of
Fermi energy for different magnitudes of strain. The dark and bright
shaded areas represent the resonant tunneling regimes for K- and
K'-valley, respectively. The transmission spectra are calculated for
given incident angle $\phi=\pi/6$. Parameters in use are $W=1$,
$L=2$ and $\beta_{ex}=2$.}\label{fg:trans_strain}
\end{figure}

The valley dependence of Dirac fermion transport through the
proposed DMB structure, of course, depends on the magnitude of
strain. Figure \ref{fg:trans_strain} shows the valley-resolved
transmission spectra for different magnitudes of strain in the
suspended region. At relatively low energy ($\epsilon<4$), there are
sharp transmission peaks at specific energies which are signatures
of resonant tunneling through the DMB structure. The transmission
peaks for different valleys are distinguishable since the energy
window of the resonant tunneling (shaded area in Fig.
\ref{fg:trans_strain})is differently determined depending on the
valley index. The sharpness of the transmission peaks leads to a
great advantage to discriminate different valleys, and provides the
possibility of valley-filtering effects. Here, notice an interesting
situation that the resonant tunneling completely disappears only for
K-valley when $\beta_{s}=\beta_{ex}$. In this case, the effective
potential profile for K-valley becomes a single rectangular barrier
[see Fig. \ref{fg:trans_angle}(d)], so that transmission probability
of Dirac fermions in K-valley are perfectly suppressed by the single
barrier rather than the resonant tunneling.

\begin{figure}
\includegraphics[width=8.5cm]{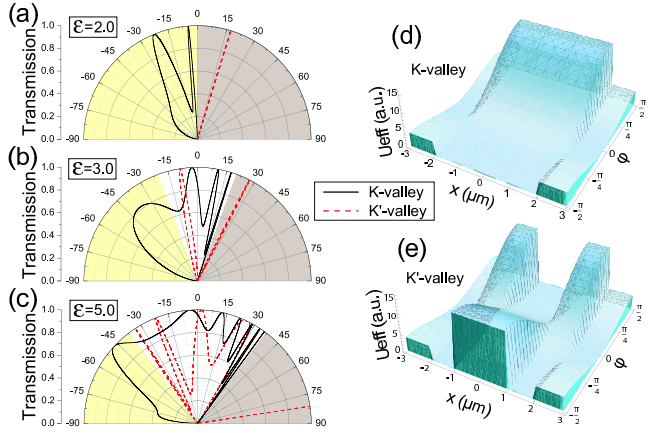}
\caption{ (a)-(c) Angle dependence of valley-resolved transmission
probability at different energies. The gray and yellow shaded areas
represent the resonant tunneling and the pseudomagnetic barrier
blockade regimes for K'-valley. Parameters in use are: $W=1$, $L=2$,
and $\beta_{ex}=\beta_{s}=2$. (d) and (e) 3D plot of the effective
potential profile as a function of the incident angle at given Fermi
energy, $\epsilon=2$, for K- and K'-valley,
respectively.}\label{fg:trans_angle}
\end{figure}

As aforementioned, the valley-dependent transport depends on the
incident angle because the effective potential varies as a function
of incident angle. The angle dependence of the valley-resolved
transmission spectra is shown in Fig. \ref{fg:trans_angle}(a)-(c).
Here, we focus on the specific case ($\beta_{ex}=\beta_{s}$) which
leads to the most dramatic change in the effective potential for
different valleys. One can see that, for Dirac fermions in
K'-valley, the resonant tunneling appears for positive incident
angles because the effective potential profile can be formed as a
double barrier. Figures \ref{fg:trans_angle}(d) and (e) display the
profile of the effective potential as a function of incident angle
for different valleys. Regardless of the incident angle, the
effective potential for K-valley is formed as a single barrier. The
effective potential for K'-valley becomes either a single or a
double barrier according to the incident angle. The formation of a
double barrier requires a condition $\phi>0$, as illustrated in Fig.
\ref{fg:trans_angle}(e), and the resonant tunneling regime
corresponds to a range of the incident angle
$\phi>\sin^{-1}{\left(1-\beta_{ex}/\epsilon\right)}$, as exhibited
in Figs. \ref{fg:trans_angle}(a)-(c). For the resonant tunneling
regimes, it is expected that Dirac fermions only in K'-valley are
allowed to pass through the DMB structure. Next, we discuss the
valley-dependent blockade of transmission probability. Figure
\ref{fg:trans_angle}(e) shows that a single pseudomagnetic barrier
is formed for negative incident angles ($\phi<0$). For K'-valley,
the probability spectra vanish for some negative angles
($\phi<-\sin^{-1}{\left(\beta_{s}/\epsilon-1\right)}$). The blockade
of Dirac fermions in K'-valley by the pseudomagnetic barrier is
exhibited as a shaded area in Figs. \ref{fg:trans_angle}(a)-(c).

\begin{figure}
\includegraphics[width=8.5cm]{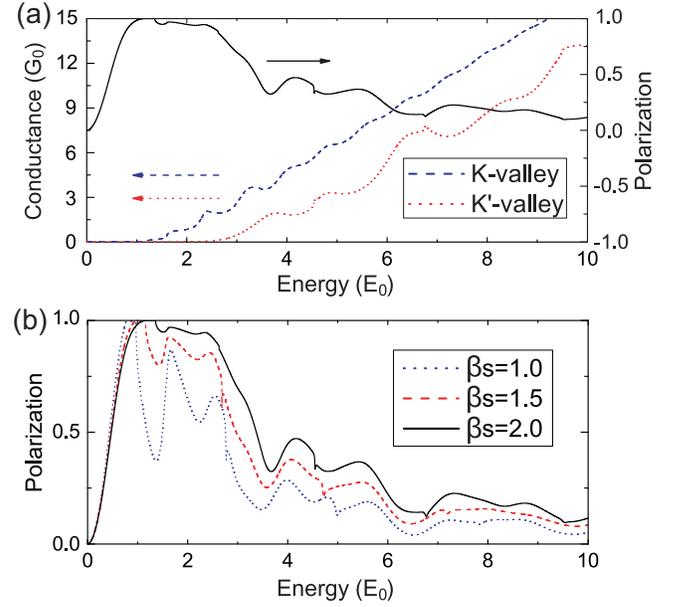}
\caption{ (a) Valley-resolved ballistic conductance and
corresponding valley polarization through the proposed DMB structure
with $\beta_{ex}=\beta_{s}=2$. The ballistic conductance is plotted
in a unit of $G_{0}=\left(e^{2}L_{y}\right)/\left(\pi
l_{B}h\right)$. (b) Strain dependence of valley polarization as a
function of Fermi energy. Parameters in use are $W=1$ and
$L=2$.}\label{fg:valleycond}
\end{figure}

The distinct angle dependence of the transmission spectra for
different valleys may lead to the valley-filtering effect. Now, we
present the valley-resolved conductance and the valley polarization
of the conductance in order to show the proposed system indeed
allows us to manipulate the valley degree of freedom. The
valley-resolved conductance spectra are exhibited in Fig.
\ref{fg:valleycond}(a) as a function of Fermi energy. Although there
are sharp resonant peaks in the transmission spectra for K-valley,
it turns out that the signature of resonant tunneling becomes very
weak in the conductance curves because of the averaging of
transmission probability over the incident angle. However, the
valley-dependent effective potential gives rise to a considerable
difference between the valley-resolved conductance. At low energy
($\epsilon<3$), the tunneling current through the DMB structure
mainly carries Dirac fermions in K-valley, yielding large valley
polarization values. Note that the valley polarization in Fig.
\ref{fg:valleycond} exhibits always positive values. The sign of
valley polarization can be changed for the opposite magnetization of
FM gates. The valley-polarization has the strain dependence as shown
in Fig. \ref{fg:valleycond}(b). As one can expect, the higher
pesudomagnetic field due to strain, the larger spin polarization of
the conductance. Here, let us make a comment on strain engineering
in practical situation. The magnitude of elastic deformation is
characterized by the ratio between the length of unstretched
graphene sheet and the vertical deformation, $h_{0}/L$. In order to
have $\beta_{s}=2$, we need $h_{0}\sim15~nm$, and the corresponding
ratio is about 1.5\%. Such value seems to be compatible with
experimental investigations because a graphene sheet can sustain
elastic deformation of the order of 20\%.\cite{R19,R20}

\section{Conclusion}

We have shown a way to produce the valley-dependent resonant
tunneling in suspended graphene as a consequence of strain effect.
We have displayed a possible model for a valley filter by using the
valley-dependent tunneling through a suspended graphene sheet
modulated by the DMB structure. The existence of strain produces
pseudomagnetic vector potentials in the suspended region, and the
resonant tunneling through the DMB structure becomes
valley-dependent because Dirac fermions in each valley differently
experience the effective potential. By tuning strain, an interesting
situation arises: the resonant tunneling can completely disappear
for Dirac fermions in K-valley. The presence of pseudomagnetic
vector potential induced by strain leads to the distinct tunneling
behavior between the valley-resolved transmission as incident angle
varies. Because of the distinct dependence on incident angle, there
appears to be significant valley polarization of the conductance in
specific energy range. The proposed valley filter based on suspended
graphene has an ability to control the valley polarization by tuning
strain in the suspended region.

In this study, we have used a minimal model with simplifications.
The simplified model indeed provide clear physical pictures to
analyze and understand the results, but one may consider some issues
further: the smooth shape of real and pseodumagnetic barriers,
effects of nonuniform external electric fields, inhomogeneous
magnetic fields at the clamped regions of suspended graphene, etc.

Lastly, we point out the possibility of spin transport in the
proposed system. When the DMB structure is asymmetric, the resonant
tunneling becomes spin-dependent and one can expect the considerable
spin polarization.\cite{R24} The spin transport through the
asymmetric DMB structure in suspended graphene provides the
manipulation of the spin degree of freedom, in addition to the
valley transport.

\begin{acknowledgments}
This work was supported by Basic Research Program through the
National Research Foundation of Korea (NRF) funded by the Ministry
of Education (2012R1A1A4A01008299).
\end{acknowledgments}

\end{document}